\documentclass[runningheads]{llncs}
\usepackage{booktabs}
\usepackage[table,xcdraw]{xcolor}
\usepackage{graphicx}
\usepackage{caption}
\usepackage{comment}
\usepackage{subcaption}
\usepackage{makecell}
\usepackage{cite}
\usepackage{siunitx}
\usepackage[labelfont={small,bf},textfont={small,it}]{caption}

\begin{document}
\title{Measuring the Consolidation of DNS and Web Hosting Providers}
%
%
\author{Synthia Wang\inst{1} \and 
Kyle MacMillan\inst{1} \and Brennan Schaffner\inst{1} \\
\and  Marshini Chetty\inst{1}\and  Nick Feamster\inst{1}}
\institute{University of Chicago, Chicago, IL 60637, USA \email{\{qiaw,macmillan,bshaffner,marshini,feamster\}@uchicago.edu}}

%
%
\authorrunning{Anonymous}

\newcommand{\brennan}[1]{ { \color{cyan}{\Large B:} #1\color{black} } }
\newcommand{\syn}[1]{ { \color{teal}{\Large S:} #1\color{black} } }
\newcommand{\mc}[1]{ { \color{red}{\Large MC:} #1\color{black} } }

\newcommand{\fixme}[1]{ { \color{red}{\bf \Large FIX ME} #1\color{red} } }

\maketitle              
\begin{sloppypar}
\begin{abstract}
Despite the Internet's continued growth, it increasingly depends on a small set of service providers to support Domain Name System (DNS) and web content hosting. This trend poses many potential threats including susceptibility to outages, failures, and potential censorship by providers. 
This paper aims to quantify consolidation in terms of popular domains' reliance on a small set of organizations for both DNS and web hosting. We highlight the extent to which a set of relatively few platforms host the authoritative name servers and web content for the top million websites. Our results show that both DNS and web hosting are concentrated, with Cloudflare and Amazon hosting over $30\%$ of the domains for both services. With the addition of Akamai, Fastly, and Google, these five organizations host $60\%$ of index pages in the Tranco top 10K, as well as the majority of external page resources. These trends are consistent across six different global vantage points, indicating that consolidation is happening globally and popular organizations can influence users' online experience across the world.

\end{abstract}
\section{Introduction}
\label{sec:introduction}

The success of the Internet can be partially attributed to its distributed
design. Indeed, distributing services over infrastructures across many parties
has contributed to the relative security, resiliency, and accessibility of the
Internet.  Over the past several years, however, control of Internet
infrastructure has begun to consolidate in fewer organizations than before.
In particular, two critical aspects of Internet service---DNS and web
hosting---were once naturally distributed but are now increasingly operated by
relatively few providers. Various organizations, including the Internet
Society, have expressed concern over the potentially negative effects of
this so-called {Internet consolidation}: {\em ``The fact that a few
corporations dominate large parts of the Internet is not news. Today, a
handful of actors play a significant role in our increasingly-connected
societies. In this context it’s important to consider what the implications of
those trends are, not only from an economic perspective but also in terms of
how they may shape the Internet in coming
years.''}~\cite{internetsociety}

These concerns over consolidation are more than a curiosity or idle concern;
rather, they have concrete, and potentially wide-ranging, ramifications. One
consequence is reduced resilience: Over the past several years, many websites
have suffered considerable outages. For example, in October 2016, a distributed
denial of service (DDoS) attack on Dyn, a major Domain Name System (DNS)
provider, resulted in outages to more than 60 major Internet sites and services,
including CNN, Etsy, GitHub, Netflix, the {\em New York Times}, Reddit, Slack,
the {\em Wall Street Journal}, Yelp, and many others~\cite{chacos_2016}.  More
recently, September 2021 alone saw several major outages. Notably, Facebook,
WhatsApp, and Instagram were unreachable for five hours after an erroneous BGP
update withdrew routing advertisements to the authoritative DNS name servers for
these services~\cite{martinho_2021}.  In late September, a large portion of
Amazon Web Services experienced a degradation that took Signal and Nest
offline~\cite{swinhoe_2021}. Although these incidents were caused by a failure within a
single organization, the magnitude of the disruptions was so pronounced
because many Internet services depended on the functionality of that single
organization.  As people become more dependent on the availability of Internet
websites and services for work, entertainment, and communication, the social and
economic costs of these incidents increases. 

The risks of consolidation also go beyond reliability. At play are issues of
Internet censorship and platform control over online speech and marketplaces.
Large content hosts such as Amazon and Cloudflare have, in the past, exercised
discretion with shutting down websites~\cite{lyons_2021, prince_2019};
Cloudflare has also prevented certain web clients from reaching websites hosted
on its platform~\cite{fingas_2021}. The consolidation of hosting on a smaller
number of platforms, particularly when any given site is hosted on {\em only
one} of the platforms, thus poses grave risks along a number of dimensions.

This process of increasing control over Internet infrastructure and services
by a small set of organizations has been defined as \textit{Internet
consolidation}, and has been defined with a relatively broad scope: {\em ``The
most visible aspects of this involve well-recognized Internet services, but it
is important to recognize that the Internet is a complex ecosystem.  There are
many underlying services whose diversity, or lack thereof, are as important as
that of, say, consumer-visible social networks.  For instance, the diversity
of cloud services, operating systems, browser engines is as important as that
of application stores or the browsers
themselves.''}~\cite{arkko2019considerations}
This expansive definition of Internet consolidation raises questions about the
{\em extent} and {\em evolution} of this phenomenon from a variety of facets.
While some previous work has considered consolidation through the lens of
DNS traffic~\cite{10.1145/3419394.3423625,hounsel2020designing}, others have
examined the economic and political implications of a few powerful companies
dominating markets on the Internet~\cite{internetsociety}.

Despite some attention to this topic, however, little is still known about how
consolidation trends are affecting the resiliency of Internet services.
Specifically, we do not have precise measures about the websites and services
that could be vulnerable to an outage of a single, particular DNS service
provider, content delivery network, or web hosting service. Of course,
answering that ``what if'' question is challenging, due to the dynamic nature
of Internet services and the nature of dependencies in complex systems such as
the Internet; yet, we can begin to get some understanding of these
vulnerabilities by studying how critical aspects of content delivery---namely,
DNS hosting and web hosting---are consolidated for popular websites and
services. 

We study several aspects of consolidation. We first explore the extent to which
a relatively small number of organizations control DNS hosting for popular
websites (Section~\ref{sec:dns}).  We find that two organizations, Amazon and
Cloudflare, are exclusively responsible for hosting the DNS name servers for
over 30\% of domains in the Tranco top 10K (i.e., 37.1\% of domains exclusively
host their DNS on one of these sites). Additionally, we find that over 89\% of
popular websites use name servers under the same primary domain, and over 75\%
of them use a single organization to host their name servers. We also study
these phenomena for web hosting (Section~\ref{sec:web}) and find similar trends:
in particular, five organizations---Cloudflare, Amazon, Akamai, Fastly, and
Google---host about 60\% of index pages in the Tranco top 10K, as well as the
majority of external page resources for these sites. 

These findings have significant implications for the design of current and
future Internet services that are resilient to both accidental
misconfiguration and overt shutdown. As others have noted, these consolidation
trends also have economic implications, particularly as they relate to issues
such as competition, barriers to entry, and permissionless
innovation~\cite{arkko2019considerations, harvard}. It is thus important
not only to report on consolidation at a particular moment in time, as we have
done in this paper, but also to track how consolidation evolves over time. To
facilitate ongoing measurements of consolidation, we have released all of our
data and measurement metrics.\footnote{The project website will be made public
upon publication of the paper.}

\section{Background and Related Work}
\label{sec:backgroun} 

This section provides background on Internet consolidation, as well as on DNS
and web hosting. We highlight particular previous work that has explored
various aspects of consolidation in these two areas. 

\subsection{The Domain Name System (DNS)} 

DNS is responsible for translating domain names into
the Internet Protocol (IP) addresses needed to communicate with the desired
network endpoint, such as a website, service, or device. Assuming the DNS record is not cached at a recursive
resolver, the client translates the domain by querying an authoritative name
server for the domain's DNS A record. If a domain name's authoritative name
servers are unreachable (and the A record is not cached), the
client cannot communicate with the network endpoint. It is common for
organizations other than the operator of the domain to host that domain's name
servers, a trend we discuss in Section~\ref{sec:dns}. As discussed in the next section, past work has
observed that increasing centralization of third-party service providers and other aspects of the DNS
can have consequences on the robustness and security of various Internet
services.

\subsubsection{DNS Consolidation.} Previous work has studied both the effects
and extent of increased DNS centralization. The previous mentioned Dyn
incident has been studied under the circumstance of DDoS attacks by Abhishta
et al.\cite{dyn_analysis_ddos}, who conclude that using multiple DNS providers
is an effective countermeasure. Bates et al. also motivate
their work by analyzing the Dyn incident and propose a metric for measuring
the market share of organizations that provide DNS resolvers~\cite{harvard}. However, since
the metric is designed based on antitrust economics literature, which is not
widely used in computer science literature, we do not apply it to our
measurements. Others have also made attempts in quantifying the consolidation
of DNS. Zembruzki et al.~\cite{dns_wild}focus on DNS route tracing and found that up to 12,000 name servers
used by websites in the Alexa top 1 million shared the same third-party
infrastructure. As for public resolvers, Radu et al. found
that the popularity of Google's public DNS resolver has increased tremendously
over time, serving as the default resolver for over 35\% of studied clients~\cite{dns_market}.
The popularity of public DNS resolvers might be explained by their lower
response times in comparison to that of local resolvers, despite the advantage
being dependent on client location~\cite{wake_dns}. To further investigate the
centralized usage of cloud providers, Moura et al. found that
more than 30\% of DNS queries to two country-code top-level domains (ccTLDs) were sent
from five large U.S. cloud providers: Google, Amazon, Microsoft, Cloudflare,
and Facebook~\cite{clouding}.


\subsection{Web Hosting and Content Delivery Networks}

Content delivery networks (CDNs) 
can provide faster delivery of content (e.g., websites) to a global
population of users. Many companies who operate CDNs also help to enhance security
by providing services such as DDoS mitigation. Websites and services who do
not rely on a CDN are subject to a variety of risks, including performance
degradation during traffic surges (``flash crowds''), weak protection against
DDoS attacks, higher Internet transit costs, and slower content delivery.
These providers not only provide sophisticated infrastructure, but also make
it easy for users to quickly establish hosting: whereas a decade ago, hosting
a website entailed a significant amount of system administration on the part
of the publisher, these CDNs also provide web hosting services that now make
this process turnkey.

\paragraph{Web Hosting Consolidation.} Prior to our work, the consolidation of
web hosting and CDNs has not been systematically studied. Recent advances in
CDN identification techniques now allow us to perform a more extensive study
of this question.  In addition to commercial CDN identification websites, Ager
at al. developed Web Content Cartography, which identify and classify CDN
infrastructures~\cite{cartography}. However, to avoid uncertainties in our
measurement, we do not explicitly measure CDNs for our dataset of domains.
Instead, we measure the hosting of page content and index pages. In many
cases, especially among the top 10K domains, CDNs are used to host the
website. Thus, by measuring content hosting, we can capture the consolidation
of CDNs in addition to other hosting options.  In a preliminary, concurrent
unpublished technical report, Moura et al. used the resource records in DNS
zone files to identify up to 200 million domains and the owners of the
autonomous systems that the domains belong to~\cite{hosting}. They found that
one-third of the domains they studied were hosted by Google, GoDaddy,
Cloudflare, and Amazon.  Our work complements and extends this study,
considering both popular domains, as well as both the DNS infrastructure
(i.e., authoritative name servers) and the hosting infrastructure that hosts
third-party, external resources of these popular websites, such as scripts. 

\section{Methods}\label{sec:methods}

In this section, we describe the methods that we used to measure two facets of
Internet consolidation: DNS hosting and web hosting.  

\paragraph{Overview.}
We study the authoritative name servers and hosting providers used by the
10,000 most popular domains from the September 2022 Tranco
rankings~\cite{tranco}. We use Tranco since it
aggregates the results of several other ranking methods to create an
accurate and more robust list.  For each domain, we
determine:
\begin{enumerate}
    \item the AS and organization hosting each of its {\bf name servers}, 
    \item the AS and organization hosting its {\bf index page}, and 
    \item the AS and organization hosting its {\bf external page resources}. 
\end{enumerate}
We define {\em organization} as the company or entity that owns the autonomous system in which the server is found. In addition, we
discuss the limitations of our approach.

\begin{figure}[t]
    \centering

    \includegraphics[width=\textwidth]{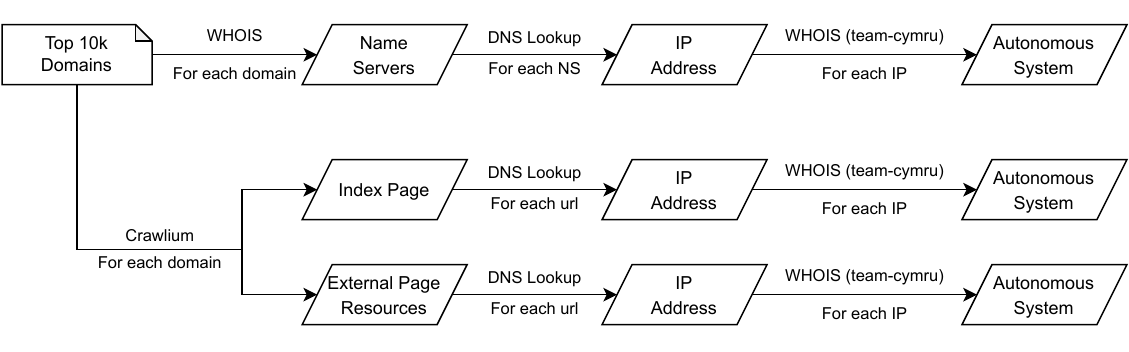}
    \caption{The measurement pipeline used to study each domain.}
    \label{fig:pipeline}
\end{figure}

\subsection{DNS Consolidation}

We identify the organization responsible for hosting the authoritative name
servers using the pipeline illustrated in Figure~\ref{fig:pipeline}. For each
domain in the top 10,000, we query the whois server referred by the ICANN whois server for its name servers. For each name server, we query the most widely used public resolvers from Google and Cloudflare, \texttt{8.8.8.8} and \texttt{1.1.1.1}, respectively, both to increase the likelihood of discovering all the IPs
associated with a given domain, and to simulate the results typical Internet
users would receive. Next, we determine the AS of each IP using
Team Cymru's~\cite{cymru_ip_2020} IP-to-AS database. 
We consider the {\em organization
name} for each AS as a distinct organization when analyzing consolidation, so
that
multiple AS numbers that share the same organization are considered as a
single organization if the organization names match.

To eliminate any biases that might result from measuring from a single
geographic location, we deploy our
measurement from six vantage points on Amazon Web Service (AWS), namely
California (US), Virginia (US), Tokyo (Japan), Mumbai (India), Frankfurt
(German), and Cape Town (Africa). We chose these vantage points to get a
global view within the AWS availability.

\subsection{Web Hosting Consolidation}
We determine the organizations responsible for hosting web content on each domain 
by examining the index page and all external page resources on the domain's 
homepage. We consider both the index page and other resources because a domain
may use different hosting providers for each. For example, Github
(\url{https://github.com}) hosts its 
own index page but uses Fastly (\url{https://www.fastly.com}) to host other assets. We scrape the homepage of 
each domain using Crawlium~\cite{crawlium}, an open-source web crawler used in prior works~\cite{haninfrastructure, bashir_longitudinal_2019}, and extract information
about each resource loaded on the page. 

\paragraph{Index pages.}
We determine the host of each domain's index page using the pipeline illustrated in
Figure~\ref{fig:pipeline}. We cannot perform a DNS lookup directly on the domain because
the domain may be different from the URL of the index page. This can occur when
an HTTP request to the domain returns a 301 or 302 response, which means that the page has been permanently, or temporarily moved to another URL that it redirects to.
For example, making a request to \texttt{nytimes.com} will respond with a redirect to \texttt{www.nytimes.com}, and \texttt{www.nytimes.com} and \texttt{nytimes.com} are translated to different IP addresses. To account for this, we use Crawlium to determine the URL of the index page by extracting
the URL of the first HTTP request to return a 200 response, which means that the request has succeeded.
We then resolve the URL of the index page and determine the organization from which it 
was loaded.

\paragraph{External page resources.}
Most modern websites rely on dynamic content fetched from third parties. Thus, in addition to the index page, we use Crawlium
to collect information about every resource that is loaded by each domain. However, 
many of these resources may not be critical to the functioning of the page. 
Because we are interested in only the resources that are absolutely required by the page, 
we filter out ads, trackers, cookies, and certain social media elements using EasyList's
block lists~\cite{easylist}. We then perform the same lookup process as we did for name servers to determine the organization 
from which each resource was loaded. 

\subsection{Data Analysis} A domain can become suddenly unavailable due to
incidents such as attacks~\cite{dyn_update_2016} and
outages~\cite{lawler_facebook_2021} that compromise Internet services that the
domain depends on. To quantify the effect of such incidents, we introduce the
notion of domains being \textit{affected} or \textit{unreachable}. For our DNS
analysis, if a domain uses a single AS to host all of its name servers, then
the domain would be \textit{unreachable} if that AS becomes unavailable. If a
domain uses an AS for at least one of its name servers, then the domain would
be \textit{affected} if the AS becomes unavailable. Similarly, for web
hosting, we consider a domain to be \textit{unreachable} when the AS that
hosts its index page becomes unavailable. Finally, if the domain has external
page resources hosted by an AS, then the domain is only \textit{affected} if
that AS becomes unavailable.




\subsection{Limitations} We determine the name servers for domains but do not
identify the name servers for all subdomains. These name servers are likely
the same, but we acknowledge that in some cases these name servers may be
different. If a subdomain uses different authoritative name servers, a
subdomain may be reachable even if the domain is not. For example, if
\texttt{nytimes.com} and \texttt{www.nytimes.com} have different name servers,
the website may still be accessible even if the name servers for
\texttt{nytimes.com} are unreachable.

Second, modern websites, especially popular ones, may use anycast DNS, which
redirects each client to a nearby server in the CDN. As a result, a website
may rely on several organizations to host web content, yet this diversity may
not be reflected if we only measure from a single vantage point---and,
naturally, our measurements are a function of the vantage points we choose.
To mitigate these effects to some degree, we perform our measurements from six
different vantage points and we have released our measurement pipeline as
open-source software to allow other researchers to continually re-evaluate our
results, over time and from a diversity of vantage points.

Finally, a website may have different hosting organization as a backup in case the primary host
is unreachable, but only rely on that organization when the primary fails. If
this situation occurs, our measurements would 
not necessarily observe the back up host and could overestimate the effects of
consolidation for failure scenarios.



\section{Findings}

In this section, we present the results of our analysis. Results from the six
vantage points are similar, so we present results measured from California;
results from other vantage points are in the Appendix \ref{sec:appendix}. We address
the following questions: 
    (1)~Which AS and AS organizations host each domain's name servers?  (Section~\ref{sec:dns})
    (2)~Which AS or AS organizations host web content (including index pages
        and external page resources)? (Section~\ref{sec:web}) 
Our results show in both cases, hosting is
dominated by a few large companies. 

\subsection{Which Organizations Provide Name Servers?}
\label{sec:dns}
We first analyze which organizations host each domains' name servers.We characterize the
potential consequences of organization outages on each domain in terms of two
metrics: (1) unreachable and (2) affected, as defined in
Section~\ref{sec:methods}. In terms of our analysis of
authoritative DNS name servers, a domain is unreachable if all of its name
servers are hosted by the same organization and said organization is down. On
the other hand, a domain is affected if at least one of the domain's name
servers is hosted by an organization and that organization is down. Based on
these definitions, the set of unreachable domains is a
subset of the set of affected domains.

\noindent \begin{minipage}[b]{0.6\textwidth}
\centering
  \includegraphics[scale=0.4]{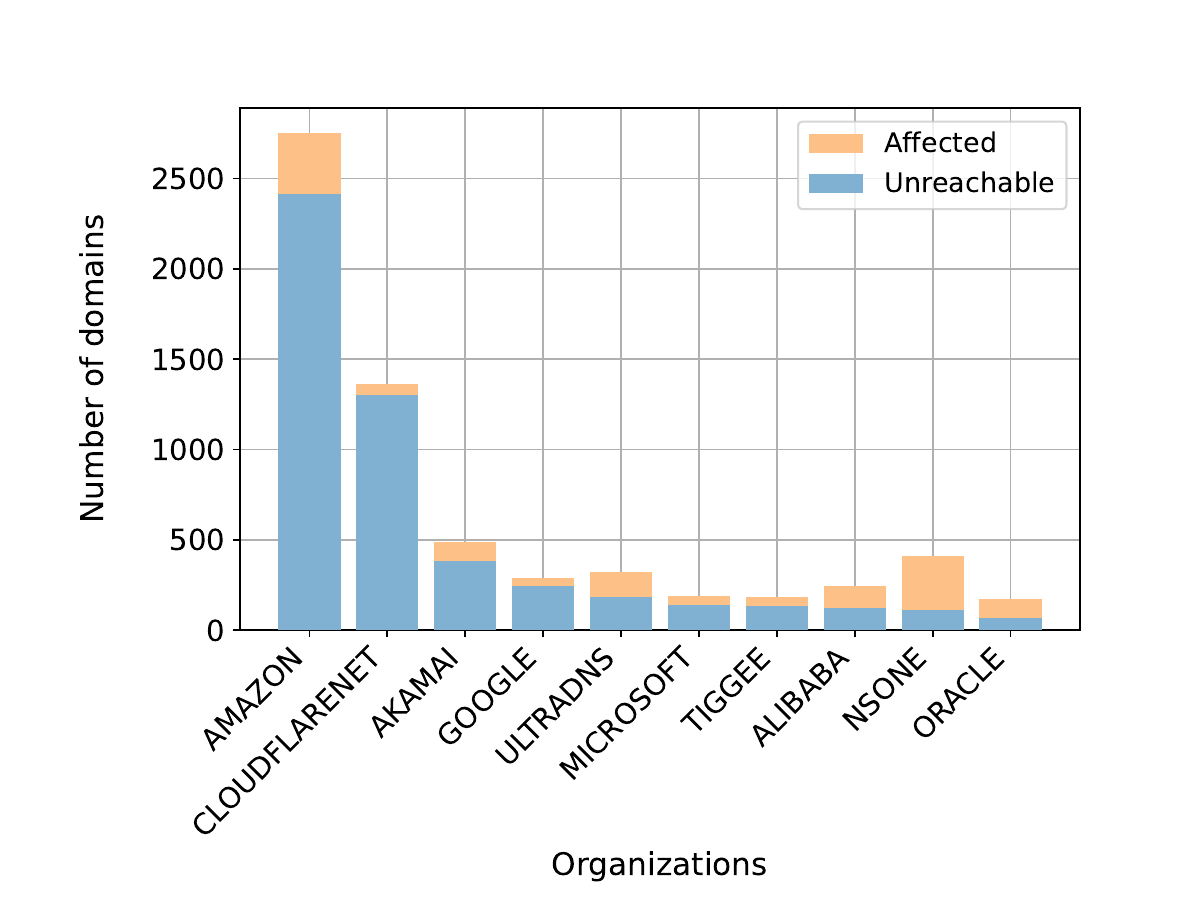}
    \setlength{\abovecaptionskip}{5pt plus 3pt minus 2pt}
		\captionof{figure}{Number of  domains that use exclusively one AS to host their name servers (unreachable) and use that AS at least partially (affected).\\}
      \label{fig:as_unreachable}
\end{minipage}
\quad
\begin{minipage}[b]{0.4\textwidth}

\centering
\scalebox{0.7}{
\begin{tabular}{lrr}
\hline
         & \% Unreachable & \% Affected                                    \\ \cline{2-3} 
Amazon & 24.1 & 27.5 \\
\rowcolor[HTML]{EFEFEF} 
Cloudflare & 13.0 & 13.6 \\
Akamai & 3.8 & 4.9 \\
\rowcolor[HTML]{EFEFEF} 
Google & 2.4 & 2.9 \\
UltraDNS & 1.8 & 3.2 \\
\rowcolor[HTML]{EFEFEF} 
Microsoft & 1.4 & 1.9 \\
TIGGEE & 1.3 & 1.8 \\
\rowcolor[HTML]{EFEFEF} 
Alibaba & 1.2 & 2.4 \\
NSONE & 1.1 & 4.1 \\
\rowcolor[HTML]{EFEFEF} 
ORACLE & 0.6 & 1.7 \\
 \hline
\rowcolor[HTML]{FFFFFF} 
Total      & 50.7          & 64.0 \\ \hline
\end{tabular}
}
\setlength{\abovecaptionskip}{57pt plus 3pt minus 2pt}
\captionof{table}{Percentages of  domains that use exclusively one AS organization to host their name servers (unreachable) and use that AS at least partially (affected).}
\label{table:as_unreachable}

\end{minipage}

\paragraph{Two organizations are responsible for hosting the name servers for over 37\% of domains.}
Amazon and Cloudflare exclusively host the name servers for $24.1\%$ and $13.0\%$ of domains, 
respectively. Table \ref{table:as_unreachable} shows the top ten most popular name server hosting
providers by the percentage of domains that would be unreachable if that provider experienced an outage.
Although these ten providers serve for over $50\%$ of domains, there is a
significant decline from Cloudflare
to the next most common organization, Akamai. This trend likely results from
the fact that Amazon and Cloudflare both offer lower-tier instant services~\cite{aws_pricing} 
and enterprise level support, whereas Akamai does not offer the lower-tier
hosting options~\cite{cdn_planet}.

\noindent \noindent\begin{minipage}[b]{0.45\textwidth}
  \centering
  \includegraphics[scale=0.3]{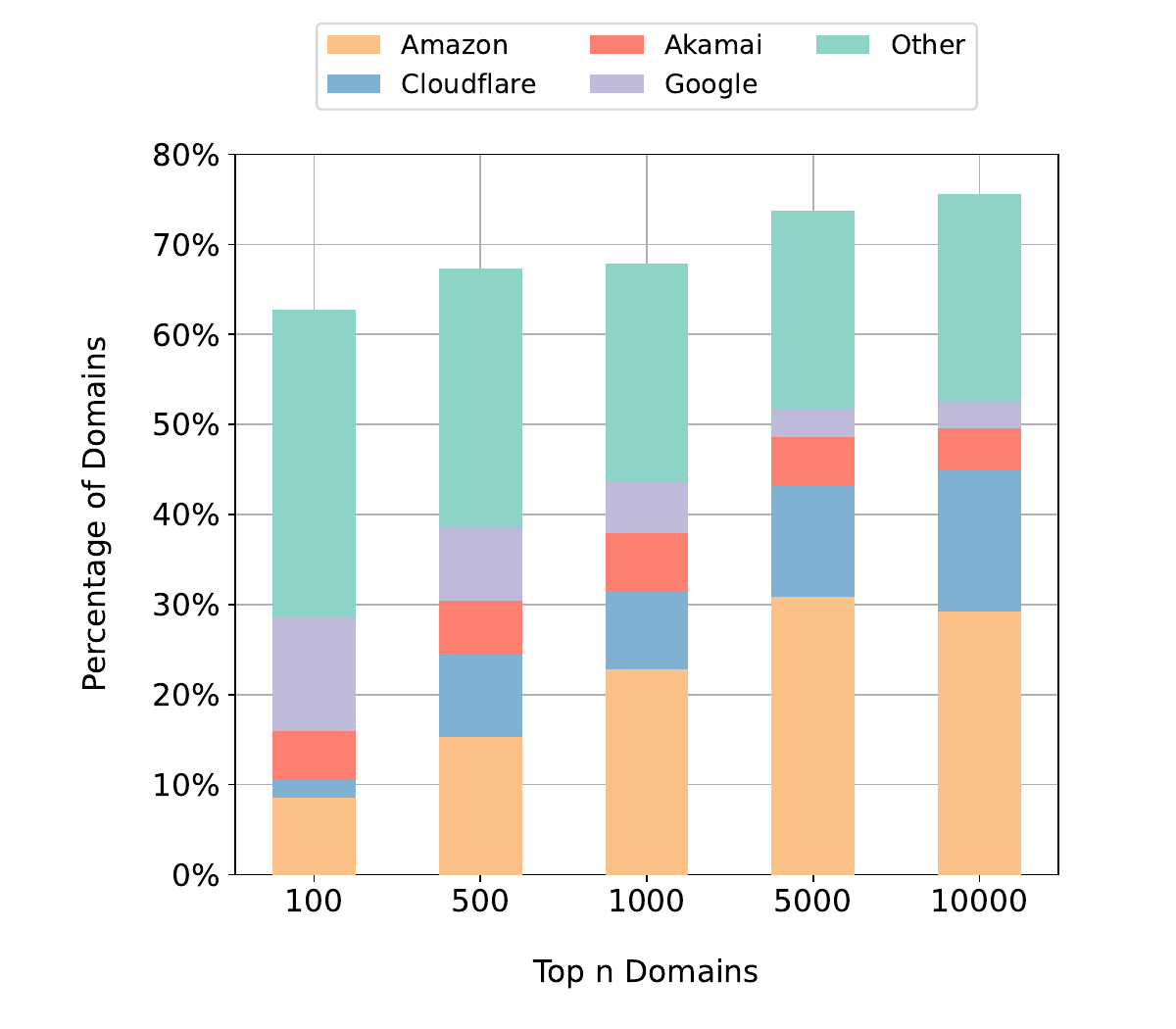}
    \setlength{\abovecaptionskip}{5pt plus 3pt minus 2pt}
  \captionof{figure}{Percentage of domains that use exclusively one \textbf{AS organization} to host their name servers (unreachable).}
  \label{fig:xas_stacked_org}
\end{minipage}
\quad
\begin{minipage}[b]{0.45\textwidth}

\centering
\includegraphics[scale=0.3]{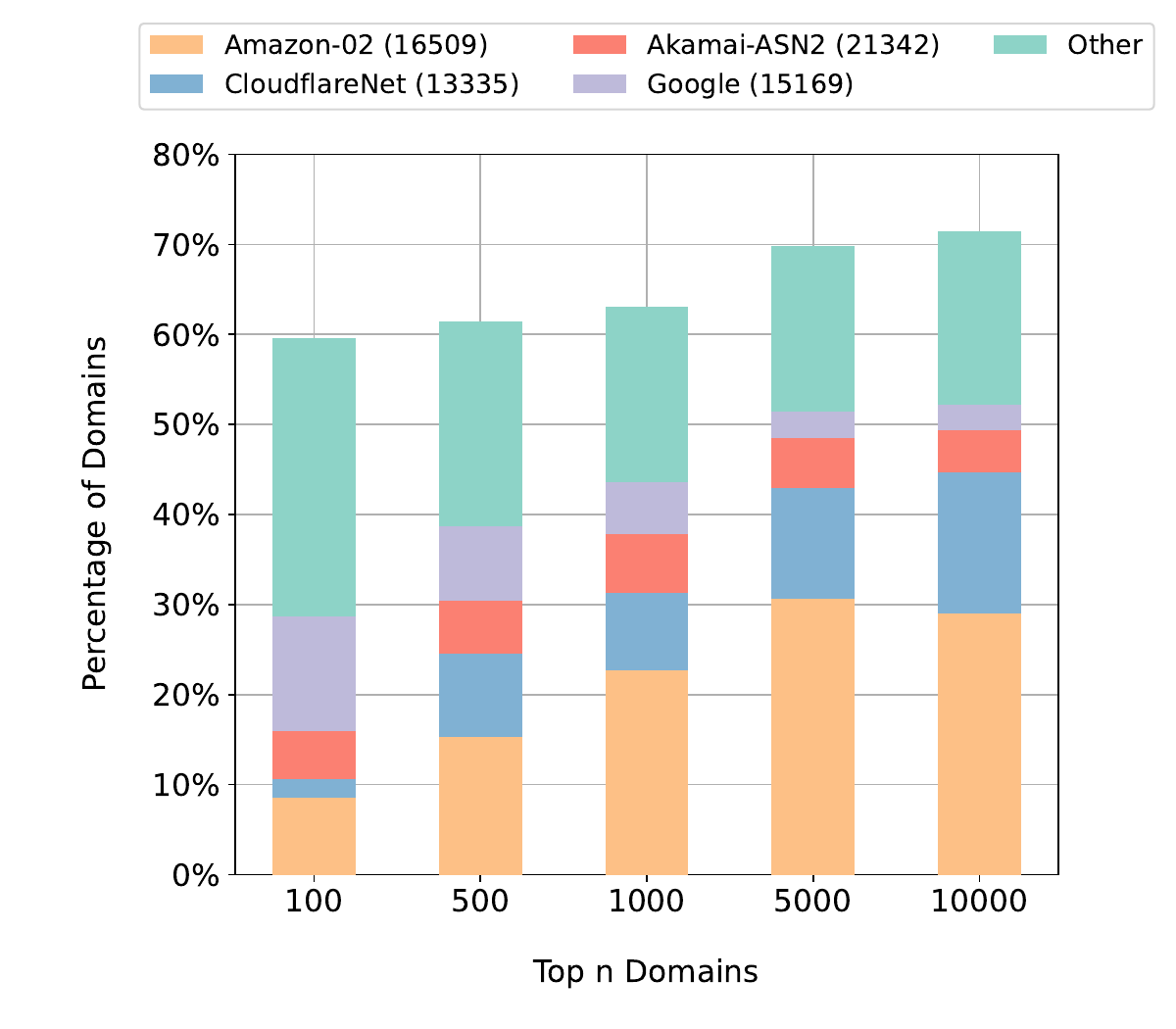}
    \setlength{\abovecaptionskip}{5pt plus 3pt minus 2pt}
  \captionof{figure}{Percentage of domains that use exclusively one \textbf{AS} to host their name servers (unreachable).}
  \label{fig:xas_stacked}
\end{minipage}

\paragraph{The use of a single organization to host all name servers is prevalent however we define the top N domains.}

Figure~\ref{fig:xas_stacked_org} shows that the percentage of domains 
that use only a given organization to host its name servers in the 
top $n$ domains is greater than $50\%$ for a wide range of possible $n$. 
In other words, both popular and less popular domains rely on a single
organization for DNS hosting. The same statement can be made when we look at ASes instead of AS organizations in Figure~\ref{fig:xas_stacked}.
However, the proportion of domains that rely on a single organization for DNS
name server hosting increases by over $20\%$ 
between the top 100 domains and the top 10,000 domains. Additionally, the 
popularity of each major organization peaks in different ranges of domains.
For example, the proportions of domains using Google peaks in the top 100 and 
decreases with $n$. This is likely because Google hosts its own 
domains but does not provide hosting services for other domains to the 
same degree as traditional hosting providers such as 
Cloudflare and Akamai. Cloudflare exhibits the opposite trend, 
hosting the lowest proportion of domains in the top 100 and increasing
with $n$. The reason for this phenomenon is unclear, although one possibility is
that Cloudflare is a less
expensive option for hosting,
and therefore more accessible, especially for the less popular domains~\cite{msv_cloudflare_2020}.

\subsection{Which Organizations Host Web Content?}
\label{sec:web}

We next study which organizations host the index page and the external page
resources for the Tranco 10K domains. Of these domains, we were able to load
the index page for 9,999; 
only 9,982 of these domains loaded
external resources that were not filtered out by our block list (e.g.,
trackers).

\begin{table}[t]
\centering
\begin{tabular}{lcrrr}
\hline
\rowcolor[HTML]{FFFFFF} 
\multicolumn{1}{c}{\cellcolor[HTML]{FFFFFF}}& \multicolumn{4}{c}{\cellcolor[HTML]{FFFFFF}Threshold} \\ \cline{2-5} 
\rowcolor[HTML]{FFFFFF} 
\multicolumn{1}{c}{{\cellcolor[HTML]{FFFFFF}}} & \hspace{10pt} ~~0\% \hspace{10pt} & \hspace{10pt} 50\% \hspace{10pt} & \hspace{10pt} 75\% \hspace{10pt} & \hspace{10pt} 90\% \hspace{10pt} \\ \cline{2-5} 
Google  & \hspace{10pt}  69.8  \hspace{10pt} & \hspace{10pt}  6.3  \hspace{10pt} & \hspace{10pt}  3.9  \hspace{10pt} & \hspace{10pt}  3.3  \hspace{10pt} \\
\rowcolor[HTML]{EFEFEF} 
Amazon  & \hspace{10pt}  56.1  \hspace{10pt} & \hspace{10pt}  13.7  \hspace{10pt} & \hspace{10pt}  6.8  \hspace{10pt} & \hspace{10pt}  3.8  \hspace{10pt} \\
Cloudflare  & \hspace{10pt}  55.5  \hspace{10pt} & \hspace{10pt}  14.3  \hspace{10pt} & \hspace{10pt}  8.9  \hspace{10pt} & \hspace{10pt}  5.6  \hspace{10pt} \\
\rowcolor[HTML]{EFEFEF} 
Akamai  & \hspace{10pt}  42.9  \hspace{10pt} & \hspace{10pt}  8.9  \hspace{10pt} & \hspace{10pt}  4.8  \hspace{10pt} & \hspace{10pt}  3.1  \hspace{10pt} \\
Fastly  & \hspace{10pt}  31.9  \hspace{10pt} & \hspace{10pt}  3.3  \hspace{10pt} & \hspace{10pt}  1.2  \hspace{10pt} & \hspace{10pt}  0.6  \hspace{10pt} \\
 \hline
\rowcolor[HTML]{FFFFFF} 
Total  & \hspace{10pt} ~-~ \hspace{10pt} & \hspace{10pt}  46.5 \hspace{10pt} & \hspace{10pt} 25.6 \hspace{10pt} & \hspace{10pt} 16.4 \hspace{10pt} \\ \hline
\end{tabular}
\setlength{\abovecaptionskip}{15pt plus 3pt minus 2pt}
\caption{Percentage of domains that fetch greater than the indicated threshold of external page
resources from a given organization. The 0\% threshold is used to count domains that load
any resources from a given organization.}
\label{exp_perc}
\end{table}

\begin{figure}[t!]
\begin{minipage}{0.45\textwidth}
  \centering
  \includegraphics[scale=0.3]{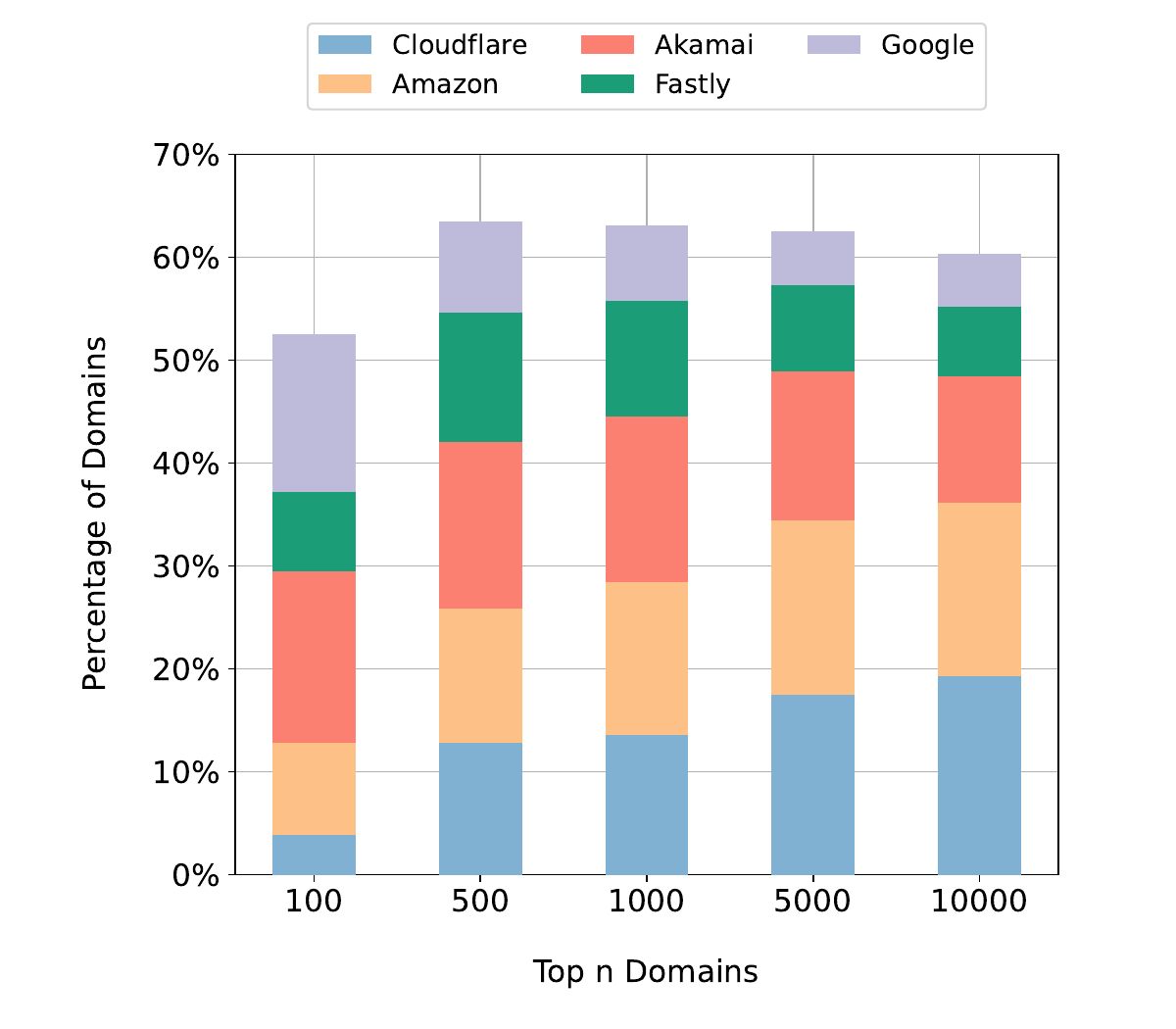}
    \setlength{\abovecaptionskip}{-5pt plus 3pt minus 2pt}
  \captionof{figure}{Percentage of  domains that use exclusively one \textbf{AS organization} to host their index page (unreachable).}
  \label{fig:index_stacked_org}
\end{minipage}
\begin{minipage}{0.45\textwidth}
  \centering
  \includegraphics[scale=0.3]{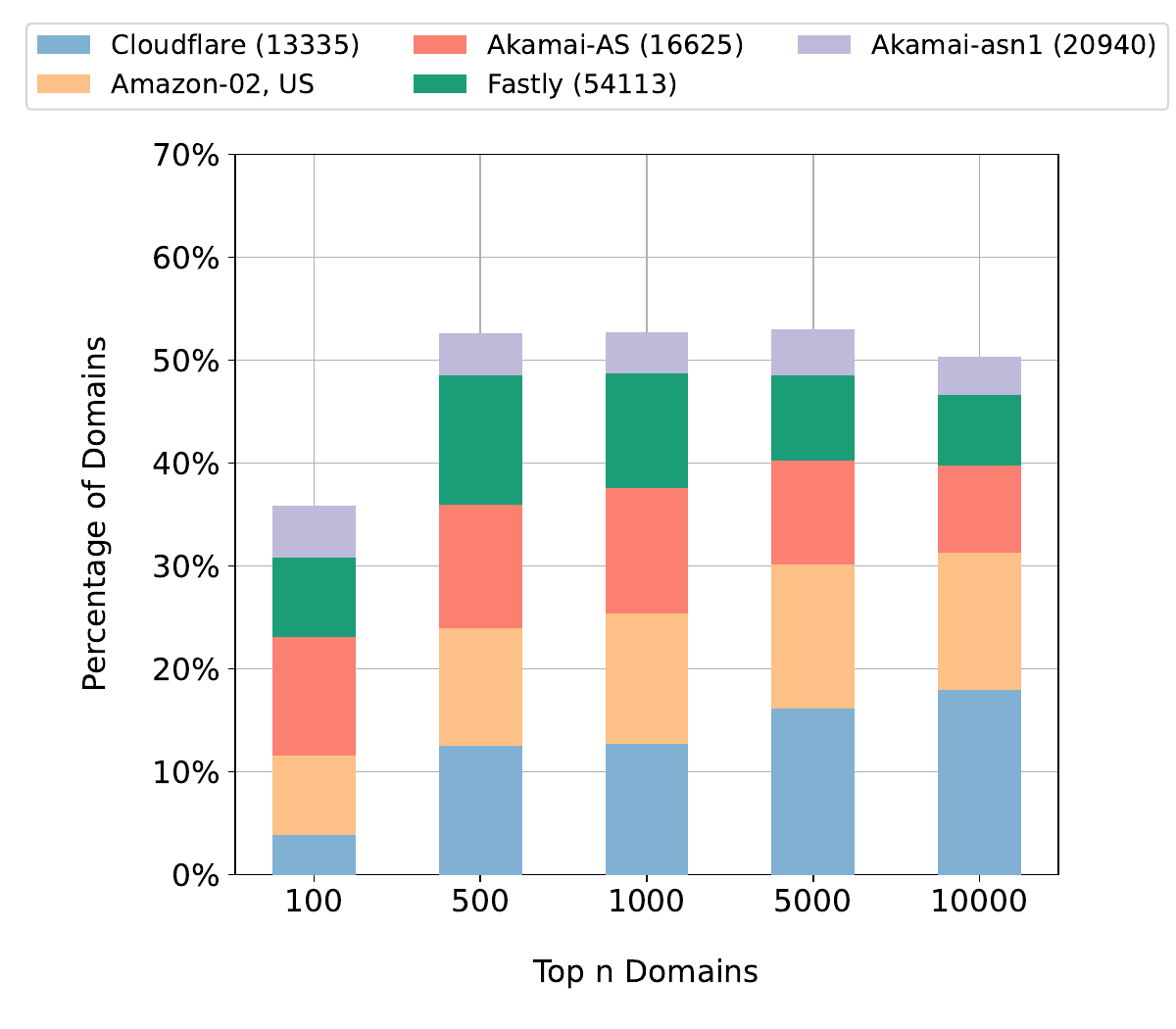}
    \setlength{\abovecaptionskip}{-5pt plus 3pt minus 2pt}
  \captionof{figure}{Percentage of  domains that use exclusively one \textbf{AS} to host their index page (unreachable).}
  \label{fig:index_stacked}
\end{minipage}
\end{figure}

\paragraph{Five organizations exclusively host the index page for a majority 
of domains.} Figure~\ref{fig:index_stacked_org} shows the percentages of index pages that
are hosted exclusively by one organization. We find that Cloudflare, Amazon, 
Akamai, Fastly, and Google exclusively host over 61\% of index pages of the
domains. As with name server hosting, 
Google's representation peaks in the top 100 domains and
Cloudflare in the top 10,000, indicating that each of the dominant organizations
may offer service packages that are attractive to or even catered towards domains
of similar popularity. With the exception of Fastly, the most popular
index page hosts are the same as the most popular name server hosts, as
presented
in Section~\ref{sec:dns}. 
In Figure~\ref{fig:index_stacked} we look at individual ASes and found that the top five ASes come from Cloudflare, Amazon, Akamai, Fastly. We see a decreased number of domain in comparison to Figure~\ref{fig:index_stacked_org}, which is caused by domains using different ASes under the same organization.

\paragraph{The same five organizations host the majority of external page 
resources.} A website may load external page resources
from many organizations, yet in many cases there is a single dominant
organization that hosts the majority of external resources for a given
webpage. We identify the websites that fetch a 
majority of resources from a single organization. To do so, we count
both the total number of resources and the number of resources
loaded from each organization. Table \ref{exp_perc} shows the five most popular sources of externally loaded content on each website's homepage
using several thresholds. For example, $14.3\%$ of domains loaded
more than $50\%$ of external page resources from Cloudflare.

More than $46\%$ of websites fetch a majority of their 
external page resources from one of the five hosting providers, namely Cloudflare, Amazon, Akamai, Fastly, and Google. Furthermore,
over $16\%$ of domains load over $90\%$ of external page resources from these
five organizations. These findings are similar to our study of 
index page hosts, in that many websites rely heavily on these five organizations for content hosting.

\section{Discussion}
\label{sec:discussion}


Our results suggest that DNS and web hosting are
concentrated around a few providers. We now discuss the implications of these
findings.

\paragraph{Resilience and security.} 
While some degree of consolidation in infrastructure can be beneficial for
efficient security upgrades and protocol implementation, domains relying on a
single organization introduces a single point of failure. As exhibited during
the Dyn and AWS outages, a large-scale DDoS attack on a single organization can
lead to many website outages. With this in mind, it is important to strike a
balance between distributing reliance across many organizations and limiting
redundancy.




Internet consolidation can also introduce the notion of ``attractive
surveillance targets''~\cite{arkko2019centralised}, where too much information
or traffic is controlled by a single entity. Companies such as Cloudflare, Amazon,
and Akamai, who host content for many domains, have access to the
network traffic to the domains, and therefore can collect more information, not
only from the domains themselves, but also from users visiting those domains.
Large amounts of user information concentrated within one system intensifies
concerns of surveillance by both the controlling company and potential
attackers.

\paragraph{Content moderation.} Market concentration can give companies the
power to selectively moderate content on platforms that they service, sometimes
enforcing stricter restrictions than are currently legally required
\cite{the_economist_2021}. As regulation continues to lag, these companies will
have greater ability to both control online content and influence future policy.
These implications are especially pertinent in light of ongoing debate on free
speech online. Multi-national companies play an even greater role in content
moderation, as they may mediate speech differentially across geo-political
lines. In this work, we explain the critical role that name server and web hosts
play in the content delivery process. Future work may continue to study how
companies can exert control at other points of this process, including at DNS
resolvers~\cite{pearce2017global}.

\paragraph{Standardizing metrics for consolidation.} Since there is no
standard metric for quantifying consolidation or guidelines on what kind of
measurements should be collected for this purpose, future work could include
proposing such metrics. Our findings on DNS hosting and content hosting only
covers limited components of the Internet infrastructure. Therefore, future
work could include more static attributes of domains such as domain
registrars. The organizations with whom the domains are (or were originally)
registered likely also illustrates a trend towards consolidation, with the
result being a decrease in the number of companies managing the domain name
registry and granting the purchasing or transferring of domain names. We also
only base our study on IPv4 address, and the IPv6 address space remains
unexplored. 


\section{Conclusion}
\label{sec:conclusion}
The consolidation of the hosting of various Internet infrastructure and
services is an ongoing trend that potentially threatens Internet resilience,
security, competition, and free and open communication. Although many
organizations, from the Internet Society to the Internet Engineering Task
Force, have expressed concern over this ongoing and evolving trend, we have,
to date, only had a limited understanding of the extent and evolution of such
consolidation. This paper aims to quantify two aspects of Internet
consolidation---authoritative DNS name server hosting and web hosting. To do
so, we analyzed the extent of consolidation of DNS authoritative name server
providers and web content hosting providers for the most popular 10,000
Internet domains, as enumerated in the Tranco 10K. 

Our analysis revealed that two organizations, Amazon and Cloudflare, are
responsible for exclusively hosting the name servers for over 40\% of domains
in the Tranco top 10K, and that only five organizations---Cloudflare, Amazon,
Akamai, Fastly, and Google---host about 60\% of index pages in the Tranco top
10K, as well as the majority of external page resources for these sites.  We
also found that more than 75\% of domains use only one AS organization for all
of its name servers. These results suggest that, in the areas of DNS name
server hosting and web hosting, consolidation is indeed significant. Our
results nonetheless represent a single snapshot of the current state of
affairs. Given the potential consequences of increasing consolidation,
continual measurements could shed more light on these trends over time and
provide insights into potential dependencies or points of vulnerability for
modern Internet services. To facilitate such measurements, we will publicly
release both our current measurements and measurement framework, and we will
continue to perform and release updated measurements periodically.

\vfill

%
%
%
\pagebreak
\bibliographystyle{splncs04}
\bibliography{refs}

\pagebreak
\setcounter{topnumber}{2}
\setcounter{bottomnumber}{2}
\setcounter{totalnumber}{4}
\renewcommand{\topfraction}{0.85}
\renewcommand{\bottomfraction}{0.85}
\renewcommand{\textfraction}{0.15}
\renewcommand{\floatpagefraction}{0.8}
\renewcommand{\textfraction}{0.1}
\setlength{\floatsep}{5pt plus 2pt minus 2pt}
\setlength{\textfloatsep}{5pt plus 2pt minus 2pt}
\setlength{\intextsep}{5pt plus 2pt minus 2pt}

\section{Appendix}
\label{sec:appendix}

This section presents data obtained from other vantage points where we deployed our measurements.
\subsection{Percentage of domains that fetch greater than the indicated threshold of external page resources from a given organization}

\begin{table}[!ht]
\begin{minipage}{0.45\textwidth}
\centering
\scalebox{0.7}{
\begin{tabular}{lcrrr}
\hline
\rowcolor[HTML]{FFFFFF} 
\multicolumn{1}{c}{\cellcolor[HTML]{FFFFFF}}& \multicolumn{4}{c}{\cellcolor[HTML]{FFFFFF}Threshold} \\ \cline{2-5} 
\rowcolor[HTML]{FFFFFF} 
\multicolumn{1}{c}{{\cellcolor[HTML]{FFFFFF}}} & \hspace{10pt} ~~0\% \hspace{10pt} & \hspace{10pt} 50\% \hspace{10pt} & \hspace{10pt} 75\% \hspace{10pt} & \hspace{10pt} 90\% \hspace{10pt} \\ \cline{2-5} 
Google  & \hspace{10pt}  69.9  \hspace{10pt} & \hspace{10pt}  6.2  \hspace{10pt} & \hspace{10pt}  3.9  \hspace{10pt} & \hspace{10pt}  3.3  \hspace{10pt} \\
\rowcolor[HTML]{EFEFEF} 
Amazon  & \hspace{10pt}  56.1  \hspace{10pt} & \hspace{10pt}  13.8  \hspace{10pt} & \hspace{10pt}  6.8  \hspace{10pt} & \hspace{10pt}  3.8  \hspace{10pt} \\
Cloudflare  & \hspace{10pt}  55.5  \hspace{10pt} & \hspace{10pt}  14.2  \hspace{10pt} & \hspace{10pt}  8.8  \hspace{10pt} & \hspace{10pt}  5.6  \hspace{10pt} \\
\rowcolor[HTML]{EFEFEF} 
Akamai  & \hspace{10pt}  42.7  \hspace{10pt} & \hspace{10pt}  8.6  \hspace{10pt} & \hspace{10pt}  4.7  \hspace{10pt} & \hspace{10pt}  3.0  \hspace{10pt} \\
Fastly  & \hspace{10pt}  32.0  \hspace{10pt} & \hspace{10pt}  3.3  \hspace{10pt} & \hspace{10pt}  1.2  \hspace{10pt} & \hspace{10pt}  0.6  \hspace{10pt} \\
\hline
\rowcolor[HTML]{FFFFFF} 
Total  & \hspace{10pt} ~-~ \hspace{10pt} & \hspace{10pt} 46.1  \hspace{10pt} & \hspace{10pt}  25.4  \hspace{10pt} & \hspace{10pt}  16.3  \hspace{10pt} \\ \hline
\end{tabular}
}
\setlength{\abovecaptionskip}{15pt plus 3pt minus 2pt}
\caption{Virginia}
\label{us_east_exp_perc}

\end{minipage}
\quad
\begin{minipage}{0.45\textwidth}
\centering
\scalebox{0.7}{
\begin{tabular}{lcrrr}
\hline
\rowcolor[HTML]{FFFFFF} 
\multicolumn{1}{c}{\cellcolor[HTML]{FFFFFF}}& \multicolumn{4}{c}{\cellcolor[HTML]{FFFFFF}Threshold} \\ \cline{2-5} 
\rowcolor[HTML]{FFFFFF} 
\multicolumn{1}{c}{{\cellcolor[HTML]{FFFFFF}}} & \hspace{10pt} ~~0\% \hspace{10pt} & \hspace{10pt} 50\% \hspace{10pt} & \hspace{10pt} 75\% \hspace{10pt} & \hspace{10pt} 90\% \hspace{10pt} \\ \cline{2-5} 
Google  & \hspace{10pt}  69.8  \hspace{10pt} & \hspace{10pt}  6.2  \hspace{10pt} & \hspace{10pt}  3.8  \hspace{10pt} & \hspace{10pt}  3.3  \hspace{10pt} \\
\rowcolor[HTML]{EFEFEF} 
Amazon  & \hspace{10pt}  56.2  \hspace{10pt} & \hspace{10pt}  14.0  \hspace{10pt} & \hspace{10pt}  7.0  \hspace{10pt} & \hspace{10pt}  3.9  \hspace{10pt} \\
Cloudflare  & \hspace{10pt}  55.5  \hspace{10pt} & \hspace{10pt}  14.3  \hspace{10pt} & \hspace{10pt}  8.8  \hspace{10pt} & \hspace{10pt}  5.6  \hspace{10pt} \\
\rowcolor[HTML]{EFEFEF} 
Akamai  & \hspace{10pt}  43.1  \hspace{10pt} & \hspace{10pt}  9.0  \hspace{10pt} & \hspace{10pt}  4.8  \hspace{10pt} & \hspace{10pt}  3.1  \hspace{10pt} \\
Fastly  & \hspace{10pt}  31.8  \hspace{10pt} & \hspace{10pt}  3.3  \hspace{10pt} & \hspace{10pt}  1.2  \hspace{10pt} & \hspace{10pt}  0.6  \hspace{10pt} \\
\hline
\rowcolor[HTML]{FFFFFF} 
Total  & \hspace{10pt} ~-~ \hspace{10pt} & \hspace{10pt}   46.8  \hspace{10pt} & \hspace{10pt}  25.6  \hspace{10pt} & \hspace{10pt}  16.5  \hspace{10pt} \\ \hline
\end{tabular}
}
\setlength{\abovecaptionskip}{15pt plus 3pt minus 2pt}
\caption{Tokyo}
\label{as_east_exp_perc}
\end{minipage}
\end{table}

\begin{table}[!ht]
\begin{minipage}{0.45\textwidth}
\centering
\scalebox{0.7}{
\begin{tabular}{lcrrr}
\hline
\rowcolor[HTML]{FFFFFF} 
\multicolumn{1}{c}{\cellcolor[HTML]{FFFFFF}}& \multicolumn{4}{c}{\cellcolor[HTML]{FFFFFF}Threshold} \\ \cline{2-5} 
\rowcolor[HTML]{FFFFFF} 
\multicolumn{1}{c}{{\cellcolor[HTML]{FFFFFF}}} & \hspace{10pt} ~~0\% \hspace{10pt} & \hspace{10pt} 50\% \hspace{10pt} & \hspace{10pt} 75\% \hspace{10pt} & \hspace{10pt} 90\% \hspace{10pt} \\ \cline{2-5} 
Google  & \hspace{10pt}  69.8  \hspace{10pt} & \hspace{10pt}  6.3  \hspace{10pt} & \hspace{10pt}  4.0  \hspace{10pt} & \hspace{10pt}  3.4  \hspace{10pt} \\
\rowcolor[HTML]{EFEFEF} 
Amazon  & \hspace{10pt}  56.1  \hspace{10pt} & \hspace{10pt}  13.9  \hspace{10pt} & \hspace{10pt}  6.9  \hspace{10pt} & \hspace{10pt}  3.9  \hspace{10pt} \\
Cloudflare  & \hspace{10pt}  55.1  \hspace{10pt} & \hspace{10pt}  14.2  \hspace{10pt} & \hspace{10pt}  8.8  \hspace{10pt} & \hspace{10pt}  5.6  \hspace{10pt} \\
\rowcolor[HTML]{EFEFEF} 
Akamai  & \hspace{10pt}  36.2  \hspace{10pt} & \hspace{10pt}  7.8  \hspace{10pt} & \hspace{10pt}  4.3  \hspace{10pt} & \hspace{10pt}  2.9  \hspace{10pt} \\
Fastly  & \hspace{10pt}  33.5  \hspace{10pt} & \hspace{10pt}  3.2  \hspace{10pt} & \hspace{10pt}  1.1  \hspace{10pt} & \hspace{10pt}  0.5  \hspace{10pt} \\
\hline
\rowcolor[HTML]{FFFFFF} 
Total  & \hspace{10pt} ~-~ \hspace{10pt} & \hspace{10pt}   45.4  \hspace{10pt} & \hspace{10pt}  25.1  \hspace{10pt} & \hspace{10pt}  16.3  \hspace{10pt} \\ \hline
\end{tabular}
}
\setlength{\abovecaptionskip}{15pt plus 3pt minus 2pt}
\caption{Mumbai}
\label{as_south_exp_perc}
\end{minipage}
\quad
\begin{minipage}{0.45\textwidth}
\centering
\scalebox{0.7}{
\begin{tabular}{lcrrr}
\hline
\rowcolor[HTML]{FFFFFF} 
\multicolumn{1}{c}{\cellcolor[HTML]{FFFFFF}}& \multicolumn{4}{c}{\cellcolor[HTML]{FFFFFF}Threshold} \\ \cline{2-5} 
\rowcolor[HTML]{FFFFFF} 
\multicolumn{1}{c}{{\cellcolor[HTML]{FFFFFF}}} & \hspace{10pt} ~~0\% \hspace{10pt} & \hspace{10pt} 50\% \hspace{10pt} & \hspace{10pt} 75\% \hspace{10pt} & \hspace{10pt} 90\% \hspace{10pt} \\ \cline{2-5} 
Google  & \hspace{10pt}  69.7  \hspace{10pt} & \hspace{10pt}  6.1  \hspace{10pt} & \hspace{10pt}  3.9  \hspace{10pt} & \hspace{10pt}  3.3  \hspace{10pt} \\
\rowcolor[HTML]{EFEFEF} 
Amazon  & \hspace{10pt}  56.1  \hspace{10pt} & \hspace{10pt}  13.7  \hspace{10pt} & \hspace{10pt}  6.7  \hspace{10pt} & \hspace{10pt}  3.9  \hspace{10pt} \\
Cloudflare  & \hspace{10pt}  55.5  \hspace{10pt} & \hspace{10pt}  14.0  \hspace{10pt} & \hspace{10pt}  8.6  \hspace{10pt} & \hspace{10pt}  5.5  \hspace{10pt} \\
\rowcolor[HTML]{EFEFEF} 
Akamai  & \hspace{10pt}  43.0  \hspace{10pt} & \hspace{10pt}  8.9  \hspace{10pt} & \hspace{10pt}  4.9  \hspace{10pt} & \hspace{10pt}  3.1  \hspace{10pt} \\
Fastly  & \hspace{10pt}  31.6  \hspace{10pt} & \hspace{10pt}  3.1  \hspace{10pt} & \hspace{10pt}  1.1  \hspace{10pt} & \hspace{10pt}  0.5  \hspace{10pt} \\
\hline
\rowcolor[HTML]{FFFFFF} 
Total  & \hspace{10pt} ~-~ \hspace{10pt} & \hspace{10pt}   45.8 \hspace{10pt} & \hspace{10pt}  25.2  \hspace{10pt} & \hspace{10pt}  16.3  \hspace{10pt} \\ \hline
\end{tabular}
}
\setlength{\abovecaptionskip}{15pt plus 3pt minus 2pt}
\caption{Frankfurt}
\label{eu_exp_perc}
\end{minipage}
\end{table}

\begin{table}[!ht]
\begin{minipage}{0.45\textwidth}
\centering
\scalebox{0.7}{
\begin{tabular}{lcrrr}
\hline
\rowcolor[HTML]{FFFFFF} 
\multicolumn{1}{c}{\cellcolor[HTML]{FFFFFF}}& \multicolumn{4}{c}{\cellcolor[HTML]{FFFFFF}Threshold} \\ \cline{2-5} 
\rowcolor[HTML]{FFFFFF} 
\multicolumn{1}{c}{{\cellcolor[HTML]{FFFFFF}}} & \hspace{10pt} ~~0\% \hspace{10pt} & \hspace{10pt} 50\% \hspace{10pt} & \hspace{10pt} 75\% \hspace{10pt} & \hspace{10pt} 90\% \hspace{10pt} \\ \cline{2-5} 
Google  & \hspace{10pt}  69.7  \hspace{10pt} & \hspace{10pt}  6.0  \hspace{10pt} & \hspace{10pt}  3.8  \hspace{10pt} & \hspace{10pt}  3.2  \hspace{10pt} \\
\rowcolor[HTML]{EFEFEF} 
Amazon  & \hspace{10pt}  56.2  \hspace{10pt} & \hspace{10pt}  14.1  \hspace{10pt} & \hspace{10pt}  7.1  \hspace{10pt} & \hspace{10pt}  4.1  \hspace{10pt} \\
Cloudflare  & \hspace{10pt}  55.5  \hspace{10pt} & \hspace{10pt}  14.1  \hspace{10pt} & \hspace{10pt}  8.8  \hspace{10pt} & \hspace{10pt}  5.6  \hspace{10pt} \\
\rowcolor[HTML]{EFEFEF} 
Akamai  & \hspace{10pt}  34.7  \hspace{10pt} & \hspace{10pt}  7.1  \hspace{10pt} & \hspace{10pt}  4.1  \hspace{10pt} & \hspace{10pt}  2.7  \hspace{10pt} \\
Fastly  & \hspace{10pt}  31.1  \hspace{10pt} & \hspace{10pt}  3.0  \hspace{10pt} & \hspace{10pt}  1.1  \hspace{10pt} & \hspace{10pt}  0.6  \hspace{10pt} \\
\hline
\rowcolor[HTML]{FFFFFF} 
Total  & \hspace{10pt} ~-~ \hspace{10pt} & \hspace{10pt}   44.3  \hspace{10pt} & \hspace{10pt}  24.9  \hspace{10pt} & \hspace{10pt}  16.2  \hspace{10pt} \\ \hline
\end{tabular}
}
\setlength{\abovecaptionskip}{15pt plus 3pt minus 2pt}
\caption{Cape Town}
\label{af_exp_perc}
\end{minipage}
\end{table}

\pagebreak
\subsection{Percentage of domains that
use exclusively one AS to host their index page (unreachable).}

\begin{figure}[!ht]
\begin{minipage}{0.45\textwidth}
  \centering
  \includegraphics[scale=0.29]{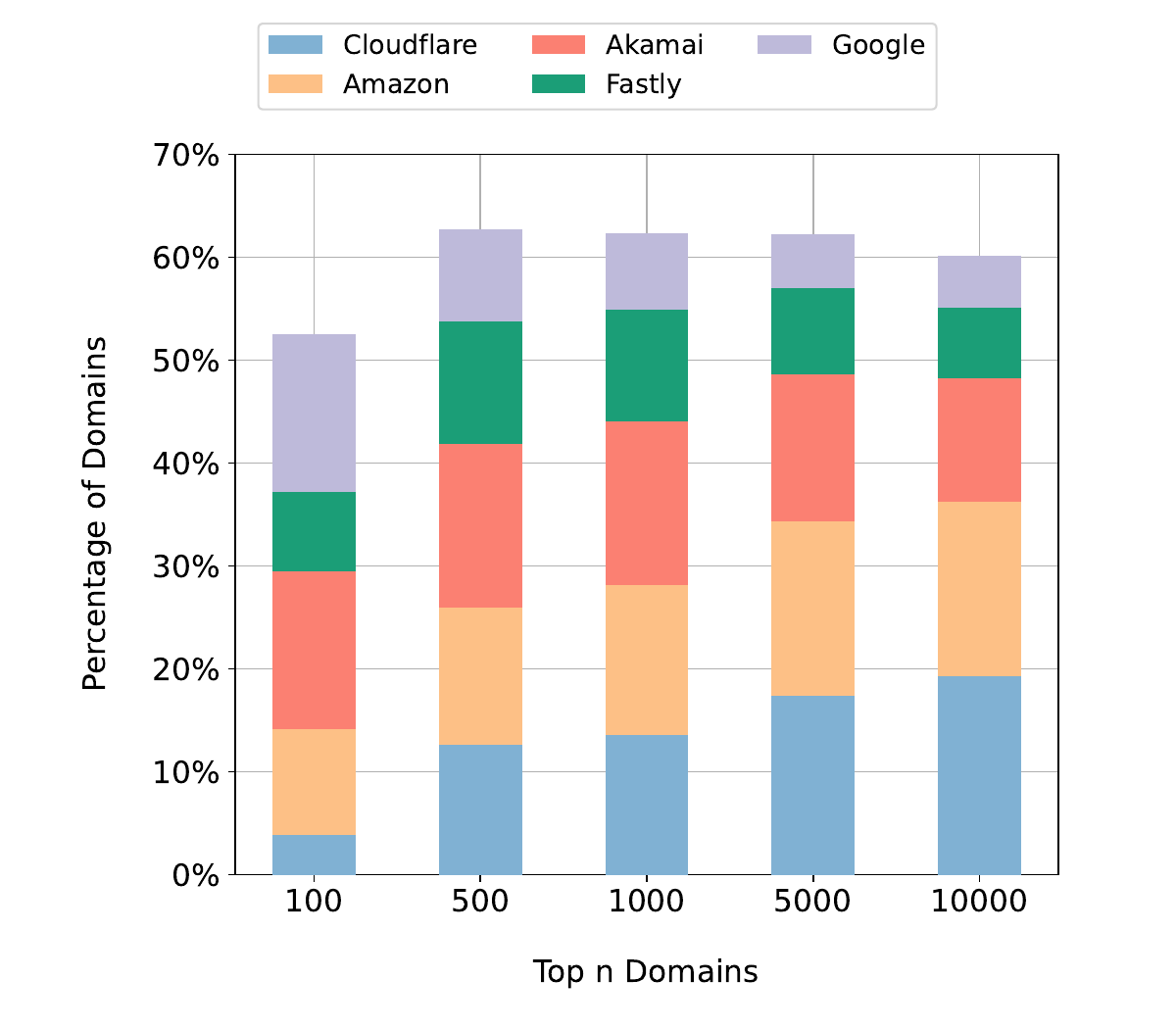}
    \setlength{\abovecaptionskip}{-5pt plus 3pt minus 2pt}
  \captionof{figure}{Virginia}
  \label{fig:us_east_index_stacked_org}
\end{minipage}
\quad
\begin{minipage}{0.45\textwidth}
  \centering
  \includegraphics[scale=0.29]{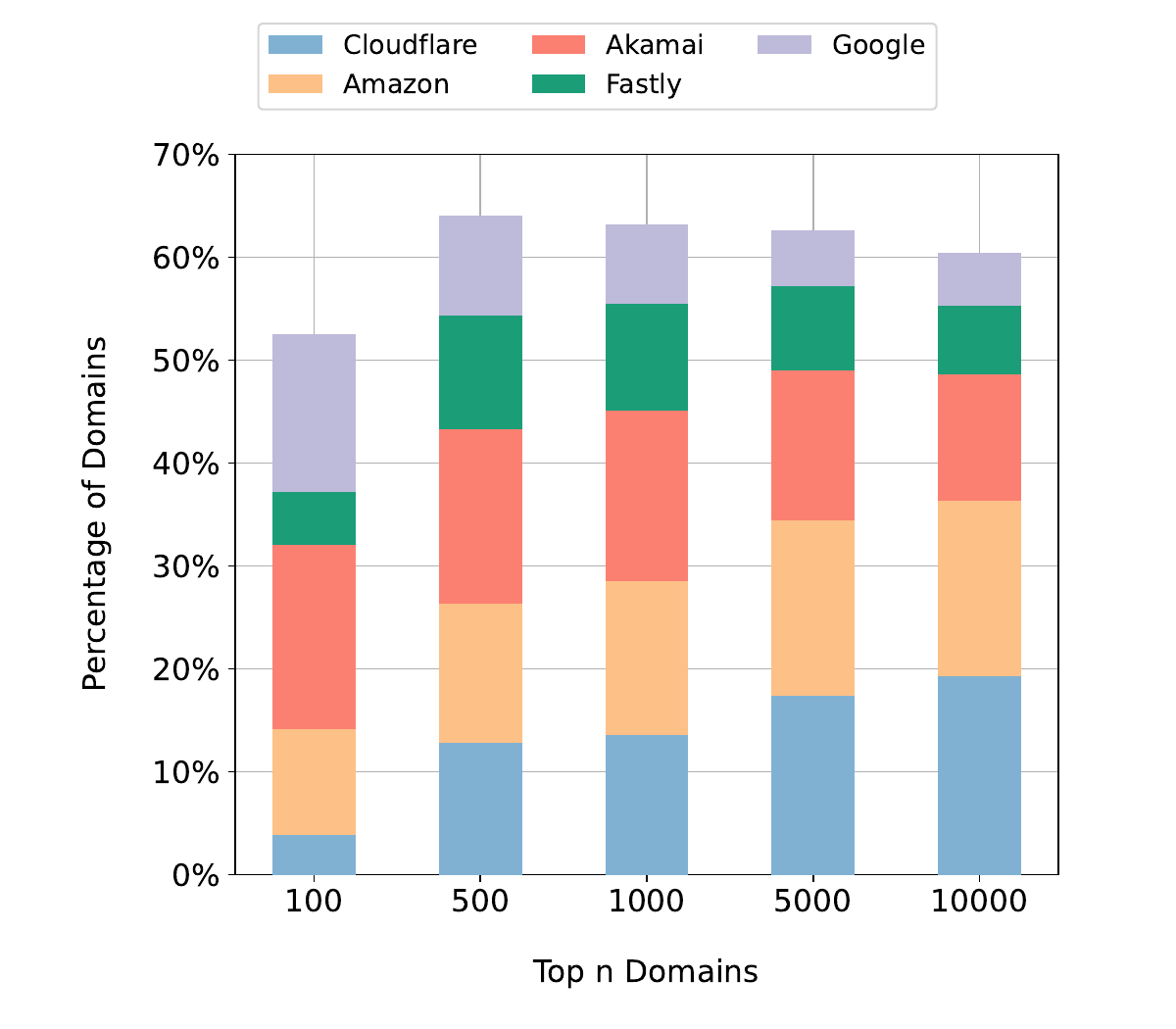}
    \setlength{\abovecaptionskip}{-5pt plus 3pt minus 2pt}
  \captionof{figure}{Tokyo}
  \label{fig:us_east_index_stacked}
\end{minipage}
\end{figure}

\begin{figure}[!ht]
\begin{minipage}{0.45\textwidth}
  \centering
  \includegraphics[scale=0.29]{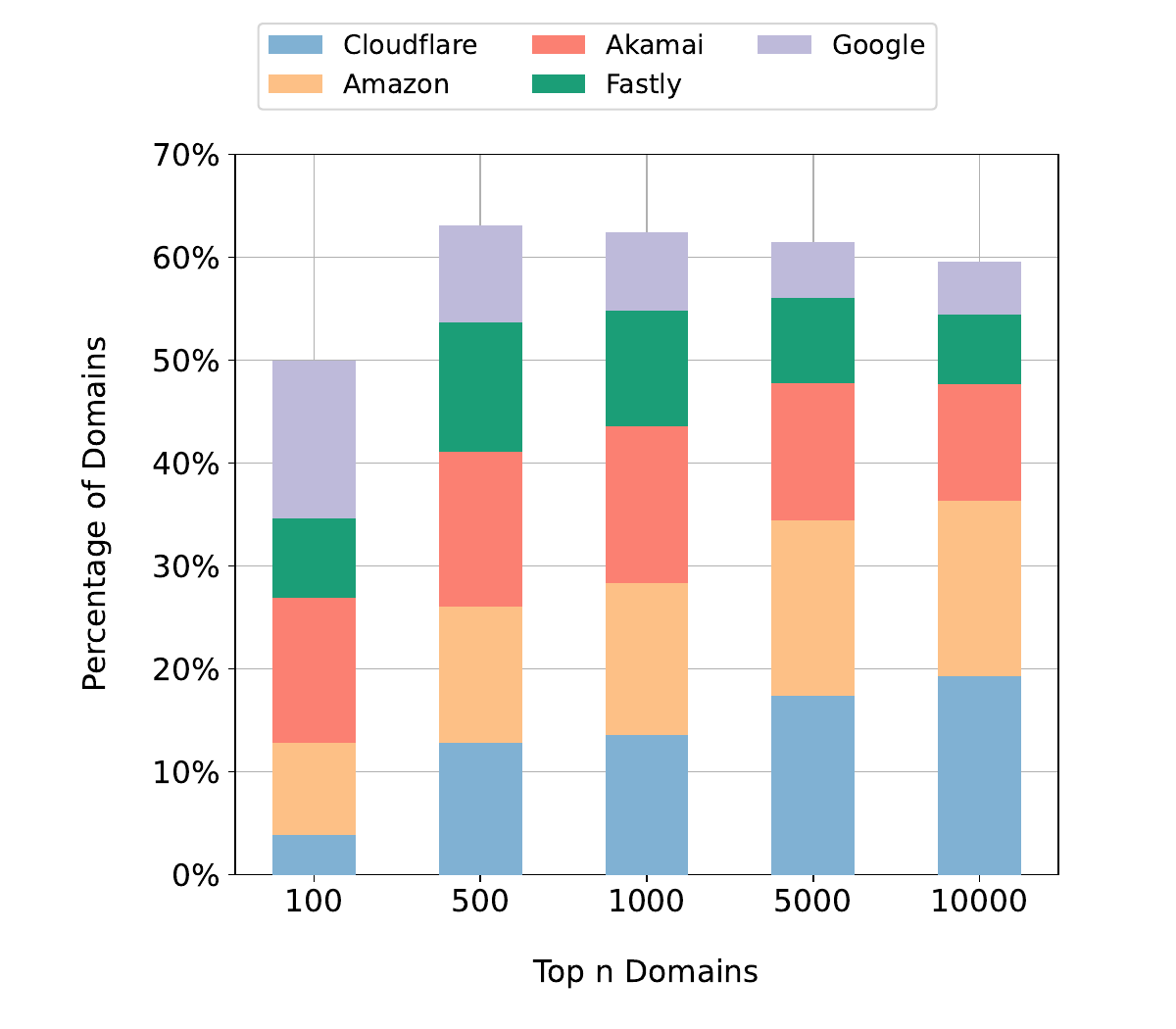}
    \setlength{\abovecaptionskip}{-5pt plus 3pt minus 2pt}
  \captionof{figure}{Mumbai}
  \label{fig:us_east_index_stacked_org}
\end{minipage}
\quad
\begin{minipage}{0.45\textwidth}
  \centering
  \includegraphics[scale=0.29]{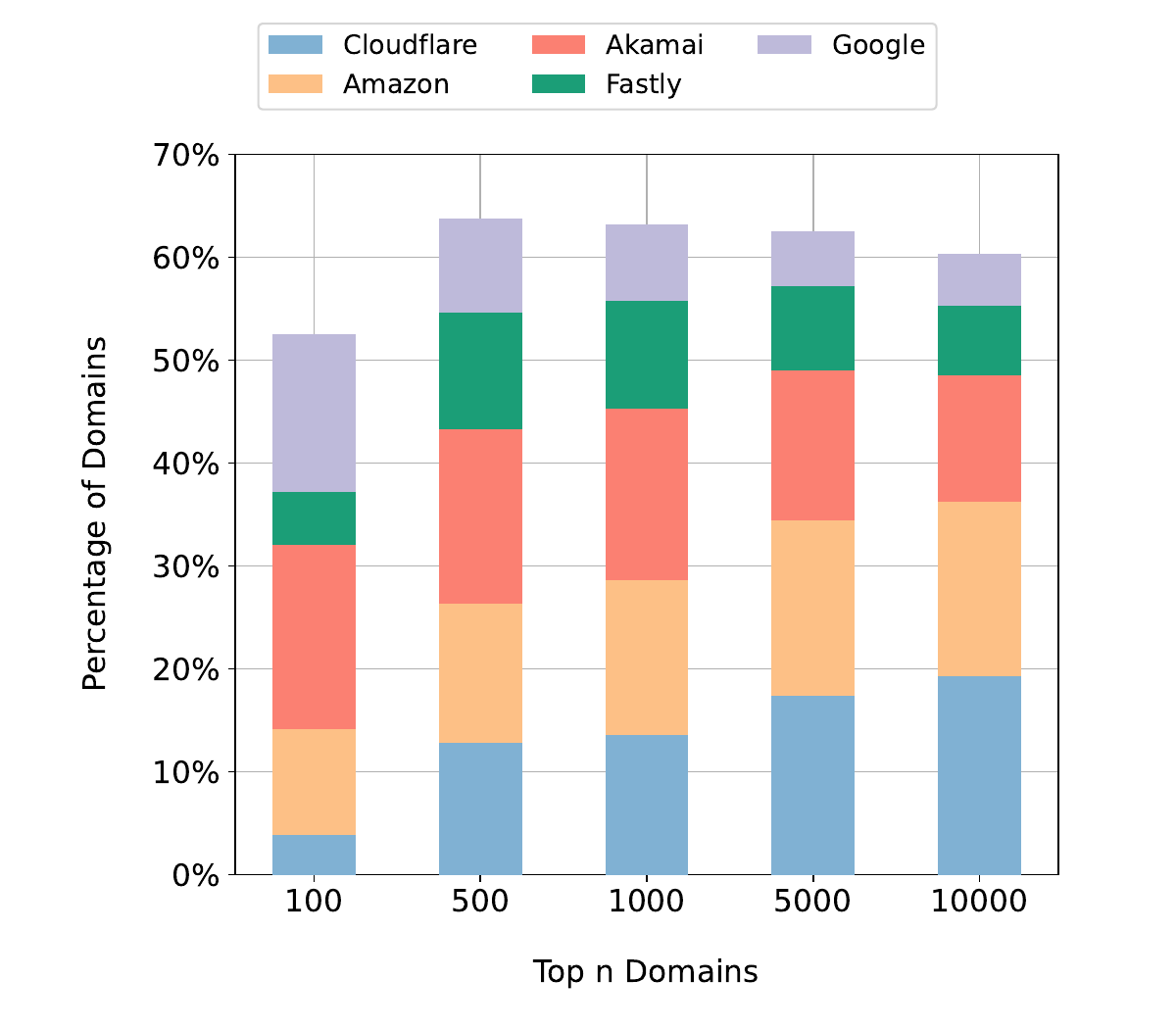}
    \setlength{\abovecaptionskip}{-5pt plus 3pt minus 2pt}
  \captionof{figure}{Frankfurt}
  \label{fig:us_east_index_stacked}
\end{minipage}
\end{figure}

\begin{figure}[!ht]
\begin{minipage}{0.45\textwidth}
  \centering
  \includegraphics[scale=0.29]{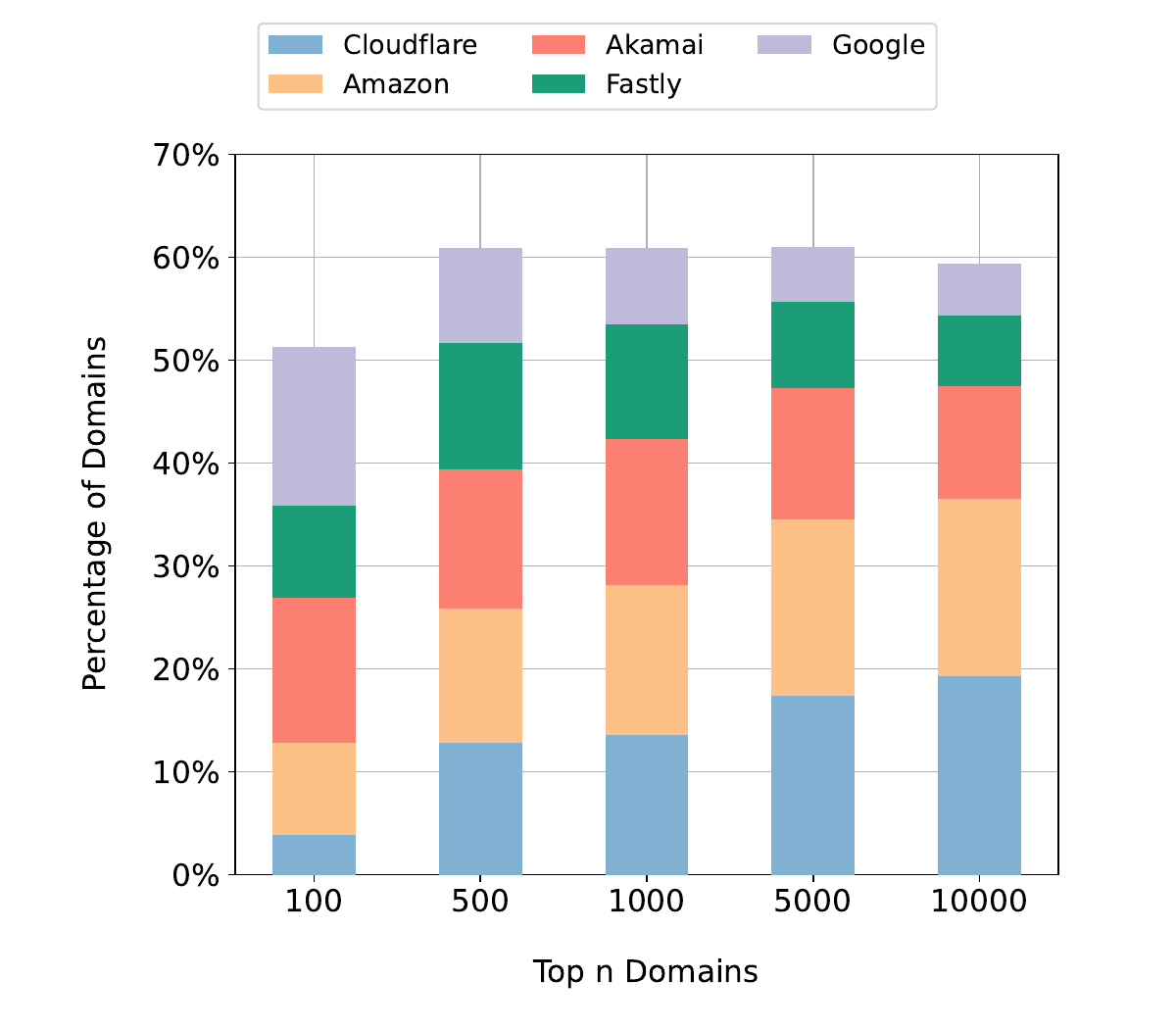}
    \setlength{\abovecaptionskip}{-5pt plus 3pt minus 2pt}
  \captionof{figure}{Cape Town}
  \label{fig:af_index_stacked}
\end{minipage}
\end{figure}

\pagebreak
\subsection{Percentage of domains that
use exclusively one AS organization to host their index page (unreachable).}

\begin{figure}[!ht]
\begin{minipage}{0.45\textwidth}
  \centering
  \includegraphics[scale=0.29]{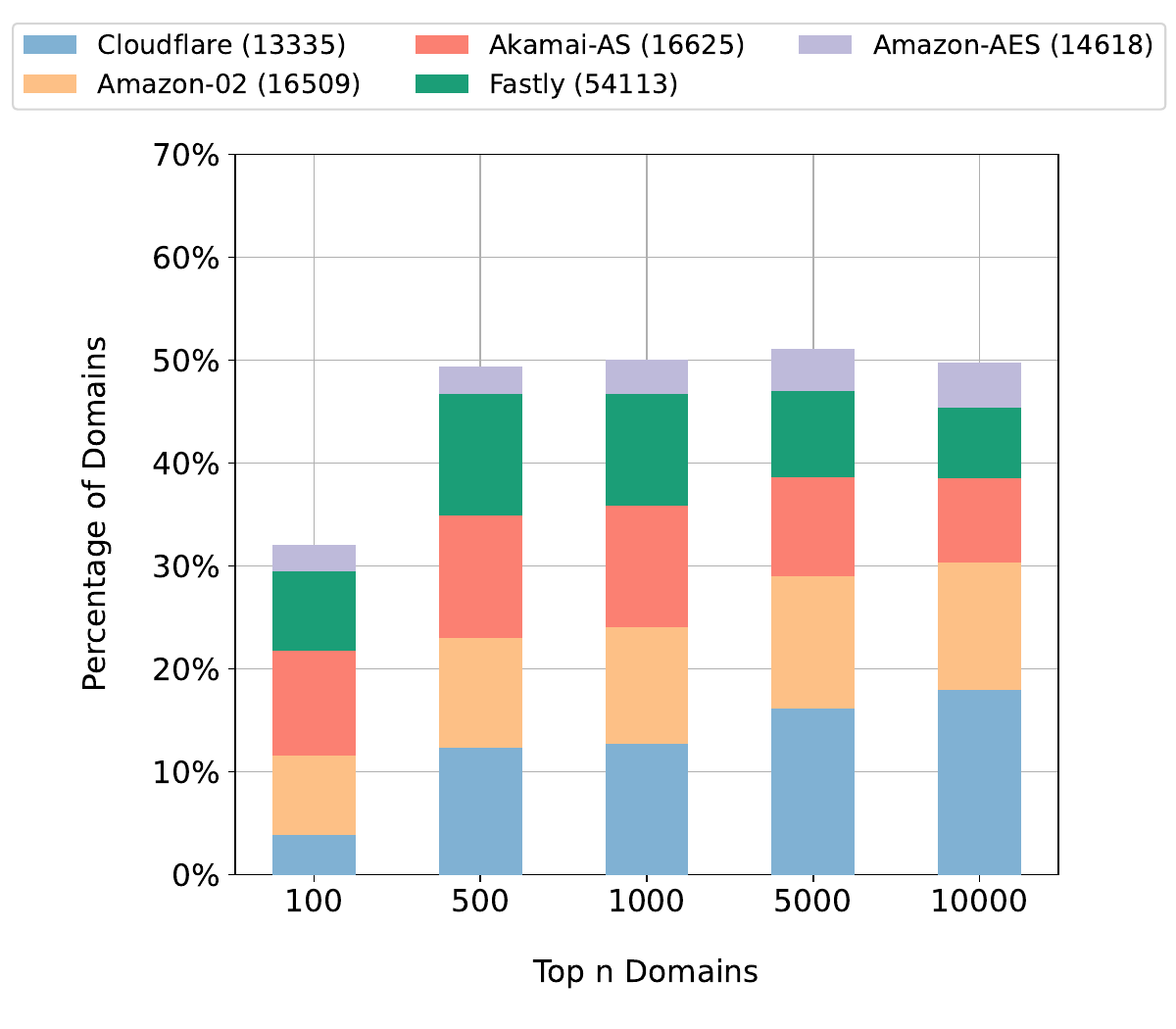}
    \setlength{\abovecaptionskip}{-5pt plus 3pt minus 2pt}
  \captionof{figure}{Virginia}
  \label{fig:us_east_index_stacked_org}
\end{minipage}
\quad
\begin{minipage}{0.45\textwidth}
  \centering
  \includegraphics[scale=0.29]{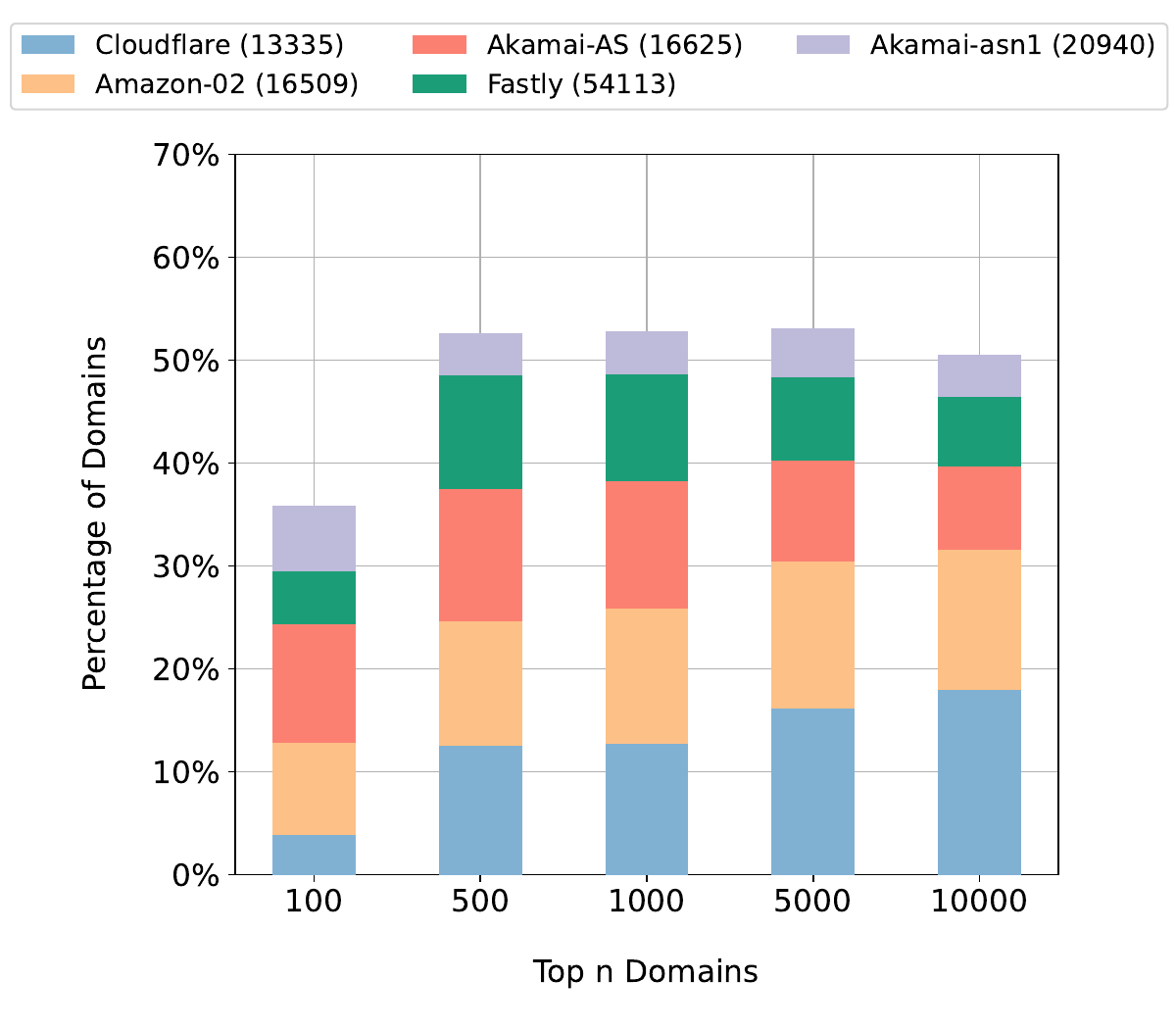}
    \setlength{\abovecaptionskip}{-5pt plus 3pt minus 2pt}
  \captionof{figure}{Tokyo}
  \label{fig:as_east_index_stacked}
\end{minipage}
\end{figure}

\begin{figure}[!ht]
\begin{minipage}{0.45\textwidth}
  \centering
  \includegraphics[scale=0.29]{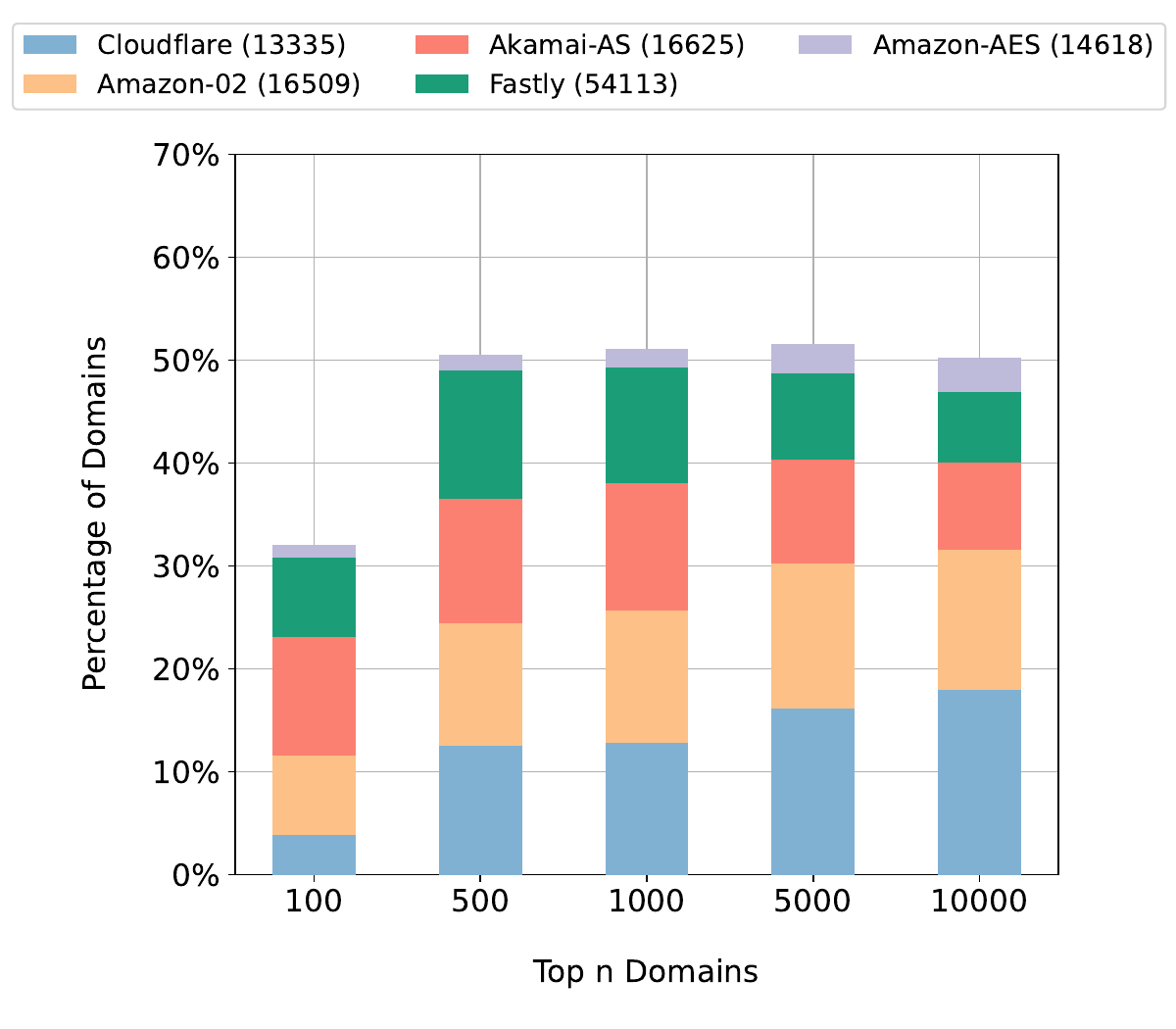}
    \setlength{\abovecaptionskip}{-5pt plus 3pt minus 2pt}
  \captionof{figure}{Mumbai}
  \label{fig:as_south_index_stacked_org}
\end{minipage}
\quad
\begin{minipage}{0.45\textwidth}
  \centering
  \includegraphics[scale=0.29]{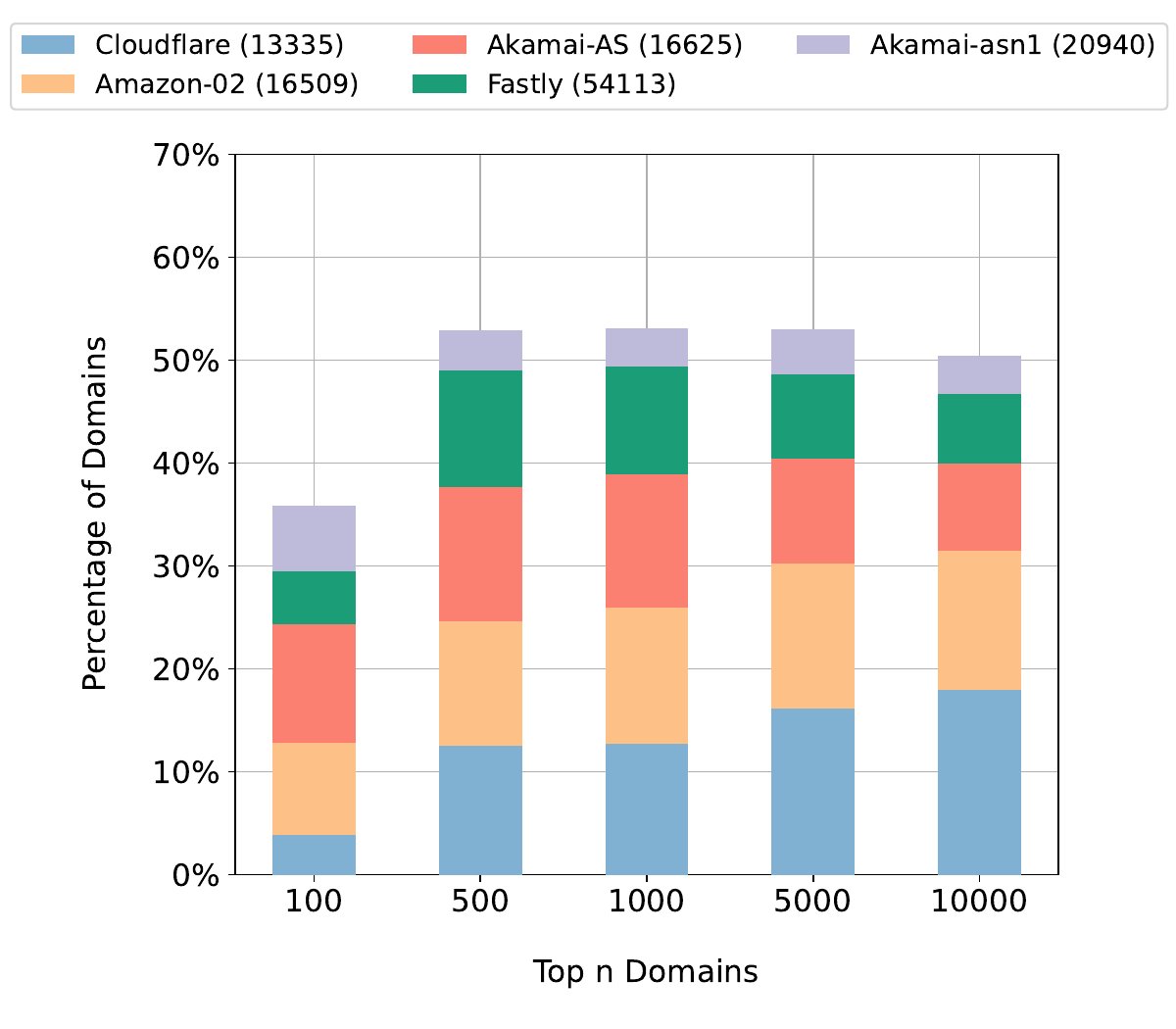}
    \setlength{\abovecaptionskip}{-5pt plus 3pt minus 2pt}
  \captionof{figure}{Frankfurt}
  \label{fig:eu_index_stacked}
\end{minipage}
\end{figure}

\begin{figure}[!ht]
\begin{minipage}{0.45\textwidth}
  \centering
  \includegraphics[scale=0.29]{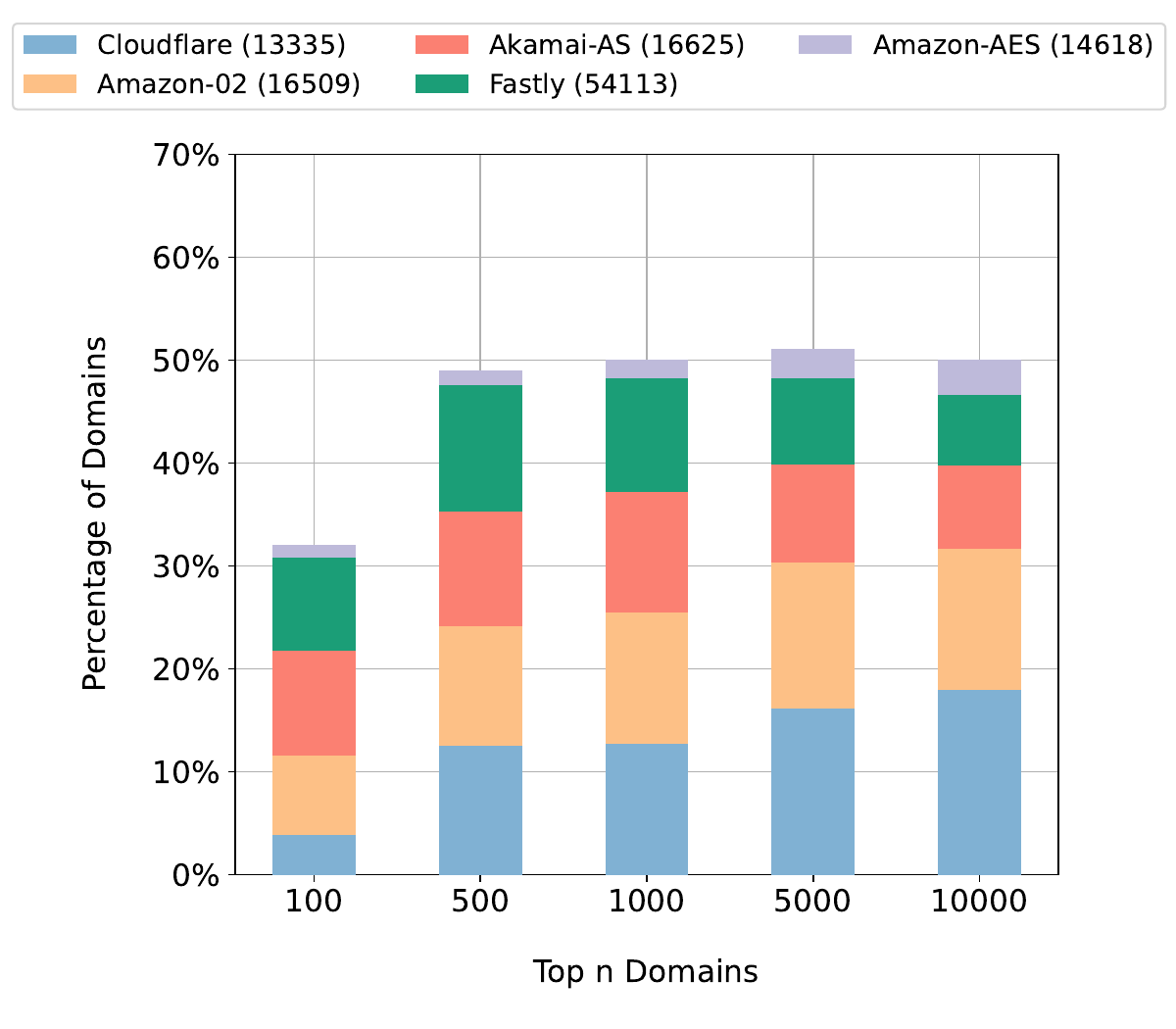}
    \setlength{\abovecaptionskip}{-5pt plus 3pt minus 2pt}
  \captionof{figure}{Cape Town}
  \label{fig:af_index_stacked}
\end{minipage}
\end{figure}

\end{sloppypar}
\end{document}